\documentclass[aps,prd,showpacs]{revtex4}
\usepackage{revsymb}

\begin{document}
\title{Schr{$\ddot{\rm{o}}$}dinger Equation and the Associated Physics in Non-Inertial Frame}
\author{Somenath Chakrabarty}
\affiliation{
Department of Physics, Visva-Bharati, Santiniketan, India  731 235, 
\\ E-mail:somenath.chakrabarty@visva-bharati.ac.in}
\begin{abstract}
In this article we have developed a formalism to obtain the solution
of Schr\"odinger equation in a non-inertial
frame. The frame is moving relative to an inertial frame with an acceleration. 
The formulation has been developed using
Lagrangian formalism as discussed in Classical Mechanics book by Landau and Lifshitz \cite{R1}. Hence
we have obtained the Hamiltonian of the nucleons. Then using the standard form of canonical
quantization rule, we have setup the Schr\"odinger equation in non-inertial frame. In the present study we have
considered only the accelerated rectilinear motion of the non-inertial frame. The rotation will be considered in
some future work based on quantum field theory \cite{JAN} (see
also \cite{DONG1,DONG2}). We therefore
drop the rotation part of the Hamiltonian in our calculation. The physically acceptable result on our work is basically
the solution obtained by Fowler and Nordheim for the emission of electrons from cold metal surface induced by strong 
electric field
\cite{R2}m known as cold field emission of electrons.
However in the present formulation,
it is the gravity which acts on mass of the particle and causes emission. The centrifugal force acts like gravity.
Hence we have got some flavor of
Hawking radiation \cite{R3} and  also Unruh effect \cite{R4} within the limited scope of our non-relativistic approach.
We have also shown that relative motion of a two body quantum mechanical system does not depend on the nature of
the frame of reference.
\end{abstract}
\maketitle
\section{Introduction}
It is well known that a reference frame at rest or moving with uniform velocity is an inertial frame. Whereas
a frame undergoing an accelerated motion with respect to some inertial frame is known as a non-inertial frame. When a
frame is rotating with uniform or non-uniform angular velocity relative to an inertial frame is also a non-inertial
frame \cite{R0,RL,BD}.
As a preamble, let me give a brief outline to
obtain the single particle Lagrangian following Landau and Lifshitz \cite{R1}. This part I believe to
be needed for the sake of completeness. Let us first consider an inertial frame $K_0$. The
Lagrangian of the particle is given by
\begin{equation}
L_0=\frac{1}{2}mv_0^2-U(\vec r)
\end{equation}
Hence using the Euler-Lagrange equation one can obtain the equation of motion of the particle. 
Next we consider a non-inertial
frame $K^\prime$ moving with a time varying rectilinear velocity $\vec V(t)$ relative to the inertial frame $K_0$.
The velocity of the electron 
in these two frames are related by the Galilean transformation $\vec v_0=\vec v^\prime +\vec
V(t)$. Hence the single particle Lagrangian in this non-inertial frame is given by
\begin{equation}
L^\prime =\frac{1}{2}m{v^\prime}^2-m\vec W.\vec r-U(\vec r)
\end{equation}
where $\vec W=d\vec V/dt$, acceleration of the frame. Hence one can obtain the equation of
motion. Let us now consider another frame $K$, whose origin coincides with that of
$K^\prime$ but rotates relative to $K^\prime$ with an angular velocity $\Omega$.
The velocity $v^\prime$ of the particle in $K^\prime$ frame is related to the velocity $v$ in
the $K$ frame by the relation $\vec v^\prime=\vec v+\vec \Omega \times \vec r$. Then the
Lagrangian of a particle in a non-inertial frame having rotational motion and also
non-uniform rectilinear motion is given by
\begin{equation}
L=\frac{1}{2}mv^2+ m\vec v.(\vec \Omega \times \vec r)+\frac{1}{2}m (\vec \Omega \times \vec
r)^2 -m\vec W.\vec r- U(\vec r)
\end{equation}
Then using the standard relation, the Hamiltonian is given by
\begin{equation}
H=\vec p.\vec v(\vec p) -L,
\end{equation}
Hence we have
\begin{equation}
H(\vec r,\vec p)=\frac{p^2}{2m}+U(\vec r)+m\vec W.\vec r -\frac{1}{2}(\vec \Omega \times
\vec r).(\vec \omega \times \vec r)
\end{equation}
Since our intention is to study the non-relativistic quantum mechanical motion of the particle 
in a non-inertial frame undergoing accelerated motion with respect to some inertial frame,
in the Hamiltonian the rotation part is
discarded. As has already been mentioned before that in a separate work we shall report the effect of rotation on 
spinor field.
Neglecting the rotational motion of the frame the Hamiltonian of the particle reduces to
\begin{equation}
H(\vec r,\vec p)=\frac{p^2}{2m}+U(\vec r)+m\vec W.\vec r
\end{equation}
where $U(\vec r)$ is some background potential.
Before we go into detail study of quantum mechanical motion of the particle in an accelerate frame, let us first discuss the
effect of acceleration of such non-inertial frame on macroscopic classical objects. It is
our common experience that when we are inside a vehicle or a train or inside an aircraft, then at
the time of acceleration of the carrier we are pushed backward. The reverse is true when the
motion is retarded. Same is true for any solid object on a non-inertial frame undergoing an
accelerated or retarded motion. Now the origin of any kind of push or pull on massive objects must be
gravity. The source (if not a human source or a mechanical source) may not be
visible or exist in reality. It is called a fictitious source. Therefore whenever there is
either acceleration or retardation a force in the opposite direction will act. We will see later that this is also
true for the quantum mechanical systems, i.e., in the subatomic world. This is one of the definitions of Principle of
Equivalence. More precisely the definition is that when a non-inertial frame undergoing accelerated / retarded
motion it is equivalent to the presence of gravity in the rest frame. However, the reverse
is not in general true \cite{RL}.
The magnitude of 
acceleration / retardation is exactly equal to the strength of gravity. Of course in
Newtonian mechanics the definition of principle of equivalence is that the inertial and the gravitational masses are
exactly same. To express it mathematically, let us write down the equation of motion of a
macroscopic object in an accelerated frame using Hamilton's equation:
\[
\frac{d\vec p}{dt}=-\frac{dH}{d\vec r}
\]
Hence the equation of motion is given by
\begin{equation}
\frac{d\vec p}{dt}=-m\vec W
\end{equation}
Since in reality a force is acting on a mass in the direction opposite to $\vec W$, we can rewrite the
above equation of motion in presence of gravity in the rest frame.  The form of the equation of motion is given by
\begin{equation}
\frac{d\vec p}{dt}=m\vec g
\end{equation}
where $\vec g$ and $\vec W$ are mutually in the opposite direction, but $\mid \vec g \mid=\mid \vec W\mid$, 
i.e., the strength of gravity and the magnitude of the acceleration are exactly equal. Hence the Newtonian form
of equivalence principle follows automatically. 

In the next section we shall investigate the quantum mechanical motion in accelerated
non-inertial frame. To the best of our knowledge this type of formalism has not been reported before.
\section{Quantum Mechanical Motion}
In quantum mechanical scenario, $H$, $\vec p$, $\vec W$ and $U(r)$ are treated as operators. Then we have
\begin{equation}
H=-\frac{\hbar^2}{2m}\nabla^2+m \vec W.\vec r+U(\vec r)
\end{equation}
Let us now consider a two body quantum mechanical system, e.g., either  deuteron or hydrogen
atom  or hydrogen molecule or hydrogen like atoms etc. Then we can recast the above Hamiltonian
into individual coordinates in the following form (indicated by the indices $1$ and $2$ for the two
components):
\begin{equation}
H=H_1+H_2
=-\frac{\hbar^2}{2m_1}\nabla_1^2+m_1\vec W.\vec r_1-
\frac{\hbar^2}{2m_2}\nabla_2^2+m_2\vec W.\vec r_2+U(\mid \vec r_1-\vec r_2\mid)
\end{equation}
Here $U(\mid \vec r_1-\vec r_2\mid)$ is the two body potential and we have neglected
the presence of any background potential. Now instead of individual coordinates, let us 
express the above Hamiltonian in terms
of relative and centre of mass coordinates, which are given by
\begin{equation}
\vec r=\vec r_1-\vec r_2 ~~{\rm{and}}~~ \vec R=\frac{m_1\vec r_1+m_2 \vec r_2}{m_1+m_2}
~~{\rm{respectively}}~~
\end{equation}
Then the two body Schr\"odinger equation is given by
\begin{eqnarray}
&&H\psi(\vec r,\vec R)=E\psi(\vec r,\vec R) ~~{\rm{or}}~~ \nonumber \\ 
&&\left [-\frac{\hbar^2}{2M}\nabla_R^2+M\vec W.\vec R-\frac{\hbar^2}{2\mu}\nabla_r^2+U(r)
\right ]\psi(\vec r,\vec R)=E\psi(\vec r,\vec R) ~~{\rm{or}}~~ \nonumber \\
&&(H_r+H_R)\psi(\vec r,\vec R)=E\psi(\vec r,\vec R)
\end{eqnarray}
where $M=m_1+m_2$, the total mass concentrated at the centre of mass and $\mu=m_1m_2/
(m_1+m_2)$, the reduced mass. Now with the separation of variables $\psi(\vec r,\vec R)= \xi(\vec
r) \phi(\vec R)$, we have
\begin{equation}
\left [-\frac{\hbar^2}{2M}\nabla_R^2+M\vec W.\vec R\right ]\phi(\vec R) =E_{cm}\phi(\vec(R)
\end{equation}
the Schr\"odinger equation in the centre of mass coordinate, and
\begin{equation}
\left [-\frac{\hbar^2}{2\mu}\nabla_r^2+U(r)\right ]\xi(\vec r)=E_{rel}\xi(\vec r)
\end{equation}
the Schr\"odinger equation in the relative coordinate. 
Obviously the relative part is independent of the acceleration of the non-inertial frame. It
is just the conventional form of Schr\"odinger equation in relative coordinate. Therefore at
this point we
can conclude that the internal activities of any two body quantum system are independent of
the nature of the frame of reference. On the other hand, the center of mass part depends on
the acceleration of the non-inertial frame. Which indicates that the centre of mass motion
will be affected by the acceleration of the frame. Here we have not considered any back
ground field. If $\theta$ is the angle between $
\vec W$ and $\vec r$, then $M\vec W.\vec r=\pm MWr\mid \cos\theta \mid$, where $+$ sign is for
the first and the fourth quadrants, whereas $-$ sign is for the second and the third
quadrants, except for $\theta=\pi/2$ and $3\pi/2$. We will see later that along these two
directions, the solutions are just the free particle 
outgoing spherical waves. In the first and fourth quadrants,
$\vec r$ is measured along the direction of acceleration $\vec W$. We call it as the forward
hemisphere. Whereas for second and third quadrants, $\vec r$ is measured along the opposite
direction of $\vec W$. We can call this direction as backward hemisphere. The problem is of
course symmetric in azimuthal coordinate. Now redefining $R^\prime =(MW\mid \cos\theta
\mid)^{1/3}R \longrightarrow R$ and $E_{cm}^*=E_{cm}/(MW\mid \cos\theta\mid )^{2/3}
\longrightarrow E$, eqn.(13) reduces to
\begin{equation}
\left [-\frac{\hbar^2}{2M}\nabla_R^2\pm R\right ]\phi(\vec R)=E\phi(\vec R)
\end{equation}
Further redefining
\[
R\longrightarrow \left (\frac{2M}{\hbar^2}\right )^{1/3}R ~~{\rm{and}}~~ E\longrightarrow
\left ( \frac{2M}{\hbar^2}\right )^{1/3}E
\]
the above differential equation may be rewritten in the following form
\begin{equation}
\left [-\nabla_R^2\pm R\right ]\phi(\vec R)=E\phi(\vec R)
\end{equation}
Now for the sake of simplicity we assume spherical symmetry, the angular momentum $\vec L=0$
and substituting
$\phi(R)=u(R)/R$, the above differential equation reduces to
\begin{equation}
\left [ -\frac{d^2}{dR^2}\pm R\right ]u(R)=Eu(R)
\end{equation}
Now instead of two body if we consider a single body, then obviously there is nothing called relative motion. As a
consequence the equation representing the relative motion (eqn.(14)) does not exist. Whereas
the centre of mass coordinate $R$ here is nothing but the single particle coordinate which coincides with the
centre of mass coordinate. We
further assuming that $U(r)$, which is now $U(R)$ is
some constant background potential. Then in the present situation the equation representing the centre of mass 
motion (eqn.(13)) is the equation for the single particle motion. Since our intention was to show that all kinds of
activities which are determined by the relative motion are independent of the characteristic of the non-inertial 
frame of reference, we started with a two body quantum system. Now we shall use eqn.(13)
satisfied by a single body and
try to extract some interesting physics with the limited scope of non-relativistic approach.  

As has already been mentioned that for $\theta=\pi/2$ or $\theta=3\pi/2$, 
the effect of acceleration on the particle motion vanishes and the
equation reduces to
\begin{equation}
\left [\frac{d^2}{dR^2}+E\right ]u(R) =0
\end{equation}
The solution is well known, which is an outgoing spherical wave, given by
\[
\phi(R)=\frac{u(R)}{R}=C \frac{\exp(-iER)}{R}
\]
where $C$ is the normalization constant. This is not the gravity induced emission. It could be some kind of
spontaneous emission, if any. 

Let us now consider the differential equation given by eqn.(17) in the backward hemisphere and re-defining
$R\longrightarrow R+E$, we have
\begin{equation}
\left [ \frac{d^2}{dR^2}+R\right ]u(R)=0
\end{equation}
To get an analytical solution we use the transformation
\[
u(R)=R^nu^\prime(R)
\]
where $n$ is an unknown constant, not necessarily an integer. Now discarding the prime symbol. 
we have
\begin{equation}
R^2\frac{d^2u}{dR^2}+2nR\frac{du}{dR}+[n(n-1)+R^3]u(R)=0
\end{equation}
To identify it with some known differential equation, let us put $R=\beta z^{2/3}$, with
$\beta$ as another constant
and $z$ is the new variable. Then we have
\begin{equation}
z^2\frac{d^2u}{dz^2} +\left (n+\frac{1}{4}\right ) \frac{4}{3}z\frac{du}{dz} +\frac{4}{9} [n(n-1)+\beta^2z^2] u(z)=0
\end{equation}
Putting $n=1/2$ and $\beta=3/2$. the above equation can be written as
\begin{equation}
z^2\frac{d^2u}{dz^2} +z\frac{du}{dz} +\left (z^2-\frac{1}{9}\right )u(z)=0
\end{equation}
Comparing with the standard form of differential equation for  Ordinary Bessel function  
\begin{equation}
z^2\frac{d^2u}{dz^2} +z\frac{du}{dz} +\left (z^2-n^2\right )u(z)=0
\end{equation}
whose solution is $J_n(z)$, the solution for eqn.(22) is $J_{1/3}(z)$. Since we expect oscillatory solution at a
large distance from the source of emission, i.e., in the asymptotic region,
we use the function $H_{1/3}^{(2)}(z)$, the Hankel function of second
kind of order ${1/3}$ instead of ordinary Bessel function. 
This is exactly the solution obtained by Fowler and Nordheim for cold field emission of
electrons from metal surface under the action of strong electric field \cite{R2}. 
However, in the
present situation, the electric field 
has to be replace by strong gravitational field. This gravitational field is acting on mass
and causes emission. This is the gravity induced emission. It is happening in a non-inertial frame
undergoing accelerated motion. The action of gravity is along the opposite direction (here it is the 
backward hemisphere). It 
has already been shown for the classical Newtonian case as well. The source of gravity is again fictitious like the classical case. This
emission process under the action of gravity may be called as the non-relativistic
version of Hawking radiation. In the relativistic picture, the creation of particle
anti-particle pairs and radiation occurs
at the event horizon of black holes. Of course the mechanism of particle production or the creation
of radiation can not be explained in the non-relativistic approach.
The Dirac vacuum, which is the source of particles, anti-particles or radiation,
does not exist in the case of Schr\"odinger equation. There is no non-relativistic counter part
of Schwinger mechanism \cite{SCH}. Therefore the solution is giving only a flavor of these quantum field theoretic phenomena.
One can also get the flavor of Unruh effect. Which is the strange field theoretic phenomenon,
in which an accelerated observer while traveling through Dirac vacuum will observe a thermal
spectrum of particle excitations. One explanation is the energy transfer from the
accelerated frame to the vacuum when the non-inertial frame interacts quantum mechanically
with the vacuum. This is also equivalent to the gravitational force acting on the particles
or anti-particles or radiation because of the accelerated motion of the frame. However, in the
non-relativistic scenario one can not distinguish a particle from its anti-particle. In such
approach, the constituents of Dirac vacuum, which are sleeping quietly are pushed out by
strong gravitational field produced by some fictitious source. The direction of force is
in the opposite direct of the
acceleration of the non-inertial frame. This is nothing but the Unruh effect and 
also equivalent to Hawking radiation. Of course this is just the qualitative explanation.

Now the solution of eqn.(17) in the forward hemisphere can also be obtained following the same
mathematical technique.
The solution is given by $J_{1/3}(iz)$. This is related to the
Modified Bessel function of first kind with argument $z$, i.e., $I_{1/3}^{(1)}(z)$. Since it
diverges asymptotically, i.e., as
$z\longrightarrow \infty$, it is therefore an unphysical solution. In reality we will never see the emission of particles
or radiation along the direction of acceleration under the action of gravity. This  is a
consequence of the  principle of
equivalence. This is also true in the classical Newtonian picture.  
\section{conclusion} In our investigation we have noticed that the relative motion of two body quantum system will
not be affected by the nature of reference frame. For a specific example, the electric quadrupole moment of deuteron,
which is quite small, will not change whether it is in inertial frame or in non-inertial frame. However. because
of motion of the frame, there will be Doppler shift of frequencies when observed from a rest frame. 
Of course this is purely classical in nature and the cause is essentially because of the relative motion between
source and the observer.

We have noticed that both in classical scenario for macroscopic objects and quantum picture for the sub-atomic
world, a force due to gravity will act on the objects. This is consistent with the principle of equivalence. Hence
we have got some flavor of Unruh effect and Hawking radiation and in consistent with the field theoretic
formulation. we found that these two effects are identical physical phenomena.

\end{document}